\documentclass[12pt]{article}
\input xy
\xyoption{all}
\input epsf.tex
\usepackage{pstricks}

\usepackage{graphicx}
\usepackage{mathrsfs}
\usepackage{cancel}
\usepackage{amssymb,amsmath}

\def\harr#1#2{\smash{\mathop{\hbox to .3in{\rightarrowfill}}
 \limits^{\scriptstyle#1}_{\scriptstyle#2}}}

\def\s2{\frac{1}{\sqrt2}}

\def\be{\begin{equation}}
\def\ee{\end{equation}}
\def\beqa{\begin{eqnarray}}
\def\eeqa{\end{eqnarray}}

\def\Dsl{\,\raise.15ex\hbox{/}\mkern-13.5mu D} 
\def\d3{d^3}

\def\IR{\mathbb{R}}

\def\IZ{\mathbb{Z}}

\newcommand{\drawsquare}[2]{\hbox{%
\rule{#2pt}{#1pt}\hskip-#2pt
\rule{#1pt}{#2pt}\hskip-#1pt
\rule[#1pt]{#1pt}{#2pt}}\rule[#1pt]{#2pt}{#2pt}\hskip-#2pt
\rule{#2pt}{#1pt}}

\newcommand{\fund}{\raisebox{-.5pt}{\drawsquare{6.5}{0.4}}}
\newcommand{\antifund}{\overline{\fund}}

\begin{document}

\begin{center}
\Large{\bf Link Invariants for Flows in Higher Dimensions  }\\
\vspace{1cm}

\large Hugo Garc\'{\i}a-Compe\'an\footnote{e-mail: {\tt
compean@fis.cinvestav.mx}}, Roberto Santos-Silva\footnote{e-mail:
{\tt rsantos@fis.cinvestav.mx}}
\\
[2mm]

{\small \em Departamento de F\'{\i}sica, Centro de
Investigaci\'on y de
Estudios Avanzados del IPN}\\
{\small\em P.O. Box 14-740, 07000 M\'exico D.F., M\'exico}\\
[4mm]

\vspace*{2cm}
\small{\bf Abstract}
\end{center}

\begin{center}
\begin{minipage}[h]{14.0cm} {Linking numbers in higher dimensions and their
generalization including gauge fields are studied in the context of $BF$ theories.
The linking numbers associated to $n$-manifolds with smooth flows generated by divergence-free $p$-vector fields, endowed with an invariant flow measure are computed in different cases. They
constitute invariants of smooth dynamical systems (for
non-singular flows) and generalizes previous results for the
3-dimensional case. In particular, they generalizes to higher dimensions the Arnold's
asymptotic Hopf invariant for the three-dimensional case. This invariant is generalized by a twisting  with
a non-abelian gauge connection. The computation of the asymptotic Jones-Witten invariants for flows is naturally extended to dimension $n=2p+1$. Finally we give a possible interpretation and implementation of these issues in the context of string theory.}
\end{minipage}
\end{center}

\bigskip
\bigskip

\date{\today}

\vspace{3cm}

\leftline{August, 2009}

\newpage
\section{Introduction}

Since their gestation topological field theories have been used to
describe topological invariants of manifolds in various dimensions.
Cohomological field theories are quantum field theories whose
states are obtained from a BRST-like symmetry. This symmetry is
usually obtained through a twist procedure starting from the set of
supercharges of an  underlying specific supersymmetric theory. In these theories physical
states are BRST-like cohomology classes of operators constructed
from the fields of the theory which, in some cases, are in correspondence with
cohomology cycles of the underlying space-time manifold. The twisted theories have
Lagrangians which depend on the background metric
however, they can be expressed as an exact BRST-like operator, and
consequently the partition function and correlation functions are
metric independent. The classical examples of these theories are the twisted ${\cal
N}=2$ Yang-Mills theory weakly coupled in four dimensions \cite{Witten:1988ze}
and \cite{monopoles} in the strong coupling limit. The underlying topological invariants are
the Donaldson and Seiberg-Witten invariants of four manifolds respectively. Another important example is the topological sigma model \cite{Witten:1988xj} giving rise to the Gromow-Witten invariants of the underlying target space.

Another kind of topological field theories are those of the Schwarz type
\cite{schwartzA}. Their Lagrangian is independent on the
background metric at the tree level but the quantization procedure
requires the introduction of a Riemannian metric in the
computation of the one-loop partition function (the
Reidemeister-Ray-Singer analytic torsion \cite{ray}) turns out to
be metric independent (up on the choice of a framing), giving a topological invariant. An example
of topological theories of this type is the Chern-Simons gauge
theory, giving rise to link invariants \cite{WJP}. In this paper
Witten found the Jones polynomial invariants of framed knots and links in terms
of correlation functions of products of oriented Wilson lines. In the
non-perturbative regime ($k \to \infty$) it was found that the theory
describes the Jones polynomial, while the perturbative one ($k \to 0$) is better written
in terms of Vassilev polynomials. (Some reviews can be found in Refs.
\cite{Atiyah:1990dn,blau-thompson,as}).

There are also inter-relations among the different
types of topological field theories. One of them is the
topological sigma model on the target space  $T^*M$ being an
hyper-Kahler manifold, and $M$ an oriented  three-manifold. The effective
theory on $M$ leads to a Chern-Simons-like theory giving rise
to perturbative link invariants known as the Rozansky-Witten invariants
\cite{Rozansky:1996bq}. There are new topological invariants
coming from the mixture of different topological field theories. Recent reviews on these topics can be found in Refs. \cite{book1,Marino:2005sj}.

Another class of theories of the Schwarz type  are the $BF$ theories
associated with higher dimensional generalizations of the
Chern-Simons gauge theories and in general of higher order
anti-symmetric tensor fields \cite{Horowitz:1989ng,Blau:1989bq}.
In these theories the link invariants in higher dimension can be
realized through the correlation functions of certain observables
of a suitable $BF$ theory
\cite{Horowitz:1989km,Blau:1989bq}. In the present paper we will follow \cite{Horowitz:1989km} in order to
extend this work to the case when there exists a multi-vector field determined by a flow on the underlying
space-time manifold.

On the other side it is very well known that topological and geometrical
methods play a crucial role in the theory of dynamical systems. Orbits in
non-singular flows can be associated with homology cycles (of
dimension one). These cycles are known as {\it asymptotic cycles}
and they were introduced some years ago by Schwartzman \cite{SA}
for the case of cycles of dimension one, i.e., they are elements
of $H_1(M,\mathbb{R})$, with $M$ being foliated. Such a ``diffuse'' cycles
are defined as the average homological placement of the periodic
orbits of the flow with respect to some invariant probability
measures. These asymptotic cycles are genuine homology cycles and
the generalization to higher dimensions was done recently in Ref.
\cite{SH}. Such generalization was achieved by using the theory of dynamical systems. In particular, the introduction of  flow boxes to define geometric currents \cite{RS,S} was very important. In particular,
the results \cite{SA} were used to carry over the
Jones-Witten polynomial invariants for flows in a dynamical system \cite{VV}. In
the process Arnold's result \cite{Arnold,Arnold-book} of the
asymptotic Hopf invariant is obtained (for recent work concerning
the asymptotic linking number, see \cite{vogel})\footnote{It is interesting to note that the Arnold's invariant coincides precisely with the {\it helicity}, which is a topological quantity arising in some physical and astrophysical processes and that remains invariant under the evolution of the system. It is precisely the behavior of the magnetic field in the plasma inside planets or stars which is described by the helicity \cite{helicityP,Arnold,Arnold-book}.}. Moreover, recently a great deal of work (basically by mathematicians)  have been done in classifying dynamical systems by using different invariants from knot theory \cite{VV,KV,badderone,baddertwo}. For a recent review on the subject see \cite{ghys}.

In the present paper we obtain a generalization of asymptotic linking numbers
in higher dimensions. They will constitute new invariants of
smooth dynamical systems. One of our main results is the fact that the correlation
function of two suitable asymptotic observables defines a higher dimensional
generalization of the average asymptotic linking number of a flow
that leaves invariant the volume form.

The paper is organized as follows. In Sec. 2 we recall the
higher dimensional description of the linking numbers mainly following
\cite{Horowitz:1989km} and \cite{Blau:1989bq}. Sec. 3 is devoted to recall the definition
of asymptotic cycles and their higher-dimensional generalization. For this we follow Ref. \cite{SH}.
We give a brief overview of the  Arnold's work concerning the
asymptotic Gauss linking number. Sections 4, 5 and 6 constitute the main part of our work. In Sec. 4 it is introduced the higher-dimensional linking number for one and two flows in the
abelian $BF$ theory (without a cosmological term). The generalized linking number with a non-abelian gauge connection is discussed in the context of one and two flows. In Sec. 5 the asymptotic Jones-Witten invariants discussed at \cite{VV} are extended to any odd ($n=2p+1$) dimension. Sec. 6 is devoted to argue on a possible relation of asymptotic invariants and string theory and at the same time it would constitutes a physical interpretation on the correlators of observables of RR fields. Finally Sec. 7 contains some concluding remarks.

\section{Overview of Linking Numbers in Quantum Field Theory}

In the present section we overview the higher dimensional linking
number. We will follows closely the work of Horowitz and Srednicki \cite{Horowitz:1989km}. The
notation and conventions are taken also from this reference.

Let $M$ be a $n$-dimensional, closed (compact and without boundary)
oriented manifold, let $U$ and $V$ be 
nonintersecting oriented submanifolds of dimension $p$ and
$p'=n-p-1$ respectively. Assume that $U$ and $V$ are homologically
trivial surfaces, that means they are boundaries of higher dimensional
surfaces. Then let $V$ be the boundary of $W$ i.e., $V =
\partial W$. We will consider also that $U$ and $W$ are going to intersect
only in a finite number of points denoted by $p_i$. We define
${\rm sign} (p_i) = 1$ if the orientation agrees with the
orientation on $M$ and ${\rm sign} (p_i) = -1$ otherwise. Then the
linking number is defined as
\begin{equation}
L(U,V) = \sum_i {\rm sign} (p_i).
\end{equation}
If $U$ and $V$ are non-intersecting circles ($p=p'=1$) embedded in
$\mathbb{R}^3$, another way of calculating the linking number is due Gauss and is given by
\begin{equation}
L(U,V) = \frac{1}{4 \pi} \int_U  dx^i \int_V dy^j \
\varepsilon_{ijk} \partial^k  |x-y|^{-1}. \label{gln}
\end{equation}

It is well known that Chern-Simons (CS) theory is used to calculate links invariants as the
Jones polynomial \cite{WJP} and the Alexander polynomial (see, for
instance \cite{Marino:2005sj}). In Refs. \cite{Horowitz:1989ng,Blau:1989bq} it was introduced a generalization
of the CS functional known as BF theory which is defined in a compact,
oriented, without boundary $n$-dimensional manifold $M$ whose
action is expressed as
\begin{equation}
\label{action1}
S_{BF} = \int_M B \wedge dC,
\end{equation}
where $B$ is a $p$-form and $C$ a $p'$ form on $M$, $d$
denotes the exterior derivative acting over forms on $M$. The equations of motion are given
by: $dB=0$ and $dC=0$. This action is invariant under diffeomorphisms
and the following gauge transformations:

\begin{equation}
\label{gauge}
B \to B + dv, \ \ \ \ \ \  C  \to C + dw,
\end{equation}
where $v$ and $w$ are $(p-1)$ and $(n-p-2)$ forms respectively.
The moduli space of the theory (the inequivalent gauge field
configurations) consist of the elements of the de Rham  cohomology
groups $H^p (M) \times H^{p'} (M)$.

The important object to compute is the two-point correlation
function of gauge invariant observables ${\cal O}_U= \int_U B$ and
${\cal O}_V= \int_V C$ given by

\begin{equation}
\label{intersectionone}
\left\langle \int_U B  \cdot \int_V C
\right\rangle = {\int {\cal D}B {\cal D}C \int_U B  \cdot \int_V C
\ e^{iS_{BF}} \over \int {\cal D}B {\cal D}C \ e^{iS_{BF}}}.
\end{equation}

In Ref. \cite{Horowitz:1989km} it was proved that this is precisely the
higher dimensional generalization of the linking number $L(U,V)$ of
two homologically trivial cycles $U$ and $V$ given by

\begin{equation}
\left\langle \int_U B \cdot \int_V C \right\rangle = i L (U,V).
\label{invone}
\end{equation}
If $p$ is odd, $n=2p+1$ and setting $C=B$, then the action becomes the
Chern-Simons functional $S=\int B \wedge d B$. In the procedure to prove (\ref{invone})  we can
choose eigenforms satisfying $\ast d B_n = \lambda_n B_n$ and consequently
we get

\begin{equation}
\left\langle \int_U B \cdot \int_V B \right\rangle = i L(U, V),
\label{invtwo}
\end{equation}
in this case $U$ and $V$ are $p$-dimensional surfaces.

For the case $M=\mathbb{R}^n$ we have that the linking number of $p$ and $p'=n-p-1$ dimensional surfaces results \cite{Wu:1988py}

\begin{equation}
L(U,V)={\Gamma({n \over 2})\over  [(2n-4) \pi^{n/2} p! ((n-p-1)!]}
\int_U d x^{j_1 \cdots j_p} \int_U d y^{j_{p+1} \cdots j_{n-1}}
\varepsilon_{j_1 \cdots j_{n-1}}^{j_n} \partial_{j_n} |x-y|^{2-n},
\label{hdlinkingnum}
\end{equation}
where $dx^{j_1 \cdots j_p} = d\sigma_1 \cdots d \sigma_{p}
J\big({x^{j_1},x^{j_2}, \cdots , x^{j_p} \over \sigma_1,\sigma_2,
\cdots ,\sigma_p} \big)$ with $J$ being the jacobian between the
coordinates $x's$ of $\mathbb{R}^n$ and the worldvolume
coordinates $\sigma's$ of $U$.  The case $U=V$ ($x=y$) is
divergent (coincident singularity) and one can have an invariant regularization procedure
\cite{WJP,Polyakov:1988md} through the choice of a framing (smooth vector field on $M$). The
asymptotic invariant that we will introduce in Sec. 4 incorporates automatically this vector field (or multi-vector field in higher dimensions) and consequently contains an invariant regularization to the self-linking
number (\ref{hdlinkingnum}) of the flow and the choice of a framing
through the choice of a $p$-vector field ${\bf X}_p= X_1 \wedge \cdots \wedge
X_p$ from the $p$-tensor product of the tangent bundle of $M$.

 \subsection{Generalized Linking Number}
 Let $A$ be a flat connection on a $G$-principal bundle $E$ over
 $M$ with $G$ a compact connected Lie group. Let us take $\mathcal{B} \in H^p(M,E)$ and
 $\mathcal{C} \in H^{p'}(M,E)$ (with $p'=n-p-1$)
 transforming in a non-trivial dual representations of $G$. Thus can take ${\cal B}$, for instance, transforming in the fundamental representation $\fund$ of $G$, while ${\cal C}$ will transform in the anti-fundamental representation $\antifund$. Consider the following action

 \begin{equation}
 \label{nonabelian}
 S = \int_M \mathcal{B} \wedge D \mathcal{C},
 \end{equation}
where $D = d + A$ is the covariant derivative and satisfies
$D^2=0$. This action is invariant under the infinitesimal gauge
transformations $ \delta\mathcal{B} = D v$ and $\delta
\mathcal{C} = D w$. The equations of motion are:  $D \mathcal{B}
= 0$ and $D \mathcal{C} = 0$. The space of solutions is the
moduli space given by $H^p(M,E) \times H^{p'}(M,E)$.
Following a similar procedure to the case without
the gauge potential (from Ref. \cite{Horowitz:1989km}) we want to extract a gauge
invariant quantity, then we must pick two surfaces $U$ and $V$ and take
a point in each one of them and a family of curves (homotopy class)
$\gamma$ that join any pair of points $x \in U$ and $y \in V$.
This family define a Wilson line, then the
correlation function of observables constructed with ${\cal B}$,
${\cal C}$ and $A$ defines a generalized linking number $L_A(U,C)$ as follows

\begin{equation}
 L_A (U,V) = -i \left \langle \int_U \mathcal{B}(x)
 \int_V P \exp \left ( \int_x^y A   \right)  \mathcal{C} (y) \right
 \rangle.
 \end{equation}
This quantity transforms as $\antifund_x\otimes (\antifund_x,\fund_y) \otimes \antifund_y$ and it is certainty an invariant of the group and consequently gauge invariant.

This expression is metric independent (because the longitudinal and zero modes do not contribute to the
integral). In this case we need that $U$ and $V$ be homologically
trivial as in the previous case, but we need one extra condition,
because the Wilson line jumps discontinuously for a non-trivial holonomy loop. Thus to ensure that
longitudinal modes do not contribute the Wilson lines are continuous
in $x$ and $y$. This statement concerning continuity is expressed in the
following definition.

A surface $U$ is \textit{holonomically} trivial if every closed
curve on $U$ has trivial holonomy.  This condition is related with
the homotopy of $U$, if every element of $\pi_1 (U)$ is homotopically
trivial in $M$ then $U$ is holonomically trivial, then the
longitudinal and zero modes not contribute to the integral. In addition there
exist holonomically trivial surfaces $W$ and $W'$ such that $\partial W = V$
and $\partial W'= U$ (because $\int_V P \exp (\int_x^y A) \mathcal{C}^0 (y) =
\int_W P \exp (\int_x^y A) D \mathcal{C}^0 (y)$ where
$D\mathcal{C}^0= 0$ and $\mathcal{C}^0$ is a zero mode). Thus we obtain the following
expression for the generalized linking number

\begin{equation}
 \label{glink}
L_A(U, V) = \sum_i {\rm sign} (p_i) {\rm Tr} P \exp
\left( \oint_{\gamma_i} A \right),
\end{equation}
with the closed curve $\gamma_i$ starting at $p_i$, following a curve
in $U$ to $u_0$, then go around $\gamma_0$ to $v_o$, and finally a curve go back to $p_i$ keeping inside $W$.

\section{Dynamics of Flows and Their Invariants}

In this section we overview briefly some definitions and mathematical
results concerning geometry and topology of dynamical systems for
future reference. Notation and conventions is taken mainly
from Refs. \cite{SA,RS,SH,dRham}. We also overview an application of these results to the
 computation of the Hopf invariant for the flow.

\subsection{Asymptotic Cycles}

One dimensional asymptotic cycles were introduced by Schwartzman
in \cite{SA}. In order to define them, we first consider a manifold
$M$ with a flow $f_t$ and take a fixed set of regular curves $\{
\gamma_{p,q} \}$ joining any pair of points $p$ and $q$ of $M$. For a given point $p$
and for any $t > 0$, we define the integral-singular
$1$-cycle $\widetilde \Gamma_{t,p}= [p,f_t(p)] \cup \gamma_{f_t (p),
p}$, where $[p, f_t(p)]$ is the oriented arc of trajectory going
from $p$ to $f_t(p)$. If we define $\Gamma_{p,t} =
\frac{1}{t} \tilde \Gamma_{t,p}$  one can prove \cite{SA} that the $\lim
_{t \to \infty} [\Gamma_{t,p}]= [\Gamma_p]$ exists and belongs to $H_1
(M, \mathbb{R})$. For every quasi-regular point $p$ of $M$, it is
independent of the metric and the connected curves $\gamma_{p,q}$ in $M$.

Now let $\mu$ be a measure on $M$ invariant under the flow and $\omega$ a closed $1$-form
in the de Rham cohomology group $H^1_{dR} ( M, \mathbb{R})$. Define the linear functional

\begin{equation}
 \Psi_\mu : H^1_{dR} ( M, \mathbb{R} ) \to \mathbb{R}, \quad \quad \quad
 \Psi_\mu ([\omega]) = \int_M \omega (X) \mu ,
\end{equation}
where $X$ is the vector field generated by the flow and $\omega$ will depend only on the
cohomology class $[\omega]$. $\Psi_\mu$ as a current \cite{dRham} on the de Rham cohomology
defines a homology cycle which can be regarded as a winding cycle
for each invariant probability ergodic measure $\mu$ of $M$.

For a higher dimensional generalization of asymptotic cycles we consider first ${\cal S}$ to be
a closed subset of a $n$-dimensional manifold $M$. A {\it partial foliation} of dimension $p$ consists of a family of $L_\alpha$ (whose dimension is such that ${\cal S}= \cup_\alpha L_\alpha$.
A collection of {\it flow boxes} on $M$ is a collection of closed disks ${\bf D}^p
\times {\bf D}^{n-p}$ (horizontal and vertical disk respectively),
whose interior cover $M$ and intersect each $L_\alpha$ in a
collection of horizontal disks $\{ {\bf D}^p \times \{y\} \}$. We
consider that the disks are smoothly embedded, such that the
tangent planes vary continuously on the flow boxes.

A $(n-p)$-dimensional submanifold $T$ of $M$ is said to be {\it transversal} if it is transversal each $L_\alpha$.
We say $T$ will be {\it small} if it is contained in a single flow box. Then a transversal
measure $\mu_T$ provides to each small transversal submanifold $T$ with a measure. We
will assume that each measure is supported on the transversal,
intersecting support of the current which is invariant under the
flow. Then we define a {\it geometrical current} as the triple $(L_\alpha,\mu_T, \nu)$, with the
entries being objects defined as above. The geometric current assigns to each point in the
support ${\cal S}$ an orientation $\nu$ of $L_\alpha$, through such point.

Suppose that $M$ is covered by a system of flow boxes $\{({\bf D}^p
\times {\bf D}^{n-p})_i\}$ endowed with partitions of unity. Then,
every $p$-form $\omega$ can be decomposed into a finite sum
$\omega = \sum_i \omega_i$, where each $\omega_i$ has his own
support in the $i$-th flow box. Now we can integrate out every
$\omega_i$ over each horizontal disk $({\bf D}^p \times \{y\})_i$
and obtain a continuous function $f_i$ over $({\bf D}^{n-p})_i$.
Thus we can take the average of this function using the
transversal measures $\mu$ to obtain a number. Therefore we
define a current given by

\begin{equation}
\langle (L_\alpha, \mu, \nu), \omega \rangle = \sum_i \int_{({\bf
D}^{n-p})_i} \mu_T(dy) \left( \int_{({\bf D}^p \times \{y\})_i}
\omega_i \right) .
\end{equation}
This current is closed in the sense of de Rham \cite{dRham}, i.e., if $\omega
= d \phi$, where $\phi$ has compact support, then $\langle
(L_\alpha, \mu, \nu), d\phi \rangle = 0$, since we can write $\phi
= \sum_i \phi_i$. Ruelle and Sullivan \cite{RS} show that this
current determines precisely an element of the $p$-th cohomology
group. In this case, if $\mu$ is invariant it does not depend of
the choice of system of flow boxes that was used. Recall that any
$(n-p)$-form $\rho$ on $M$ determines a $p$-dimensional current by
Poincar\'e duality $\langle \rho, \omega \rangle = \int_M \omega
\wedge \rho$.

Now consider an example of geometrical current. Let $\mu$ be an invariant volume
form and ${\bf X}_p$ is a $p$-vector field nowhere vanishing on $M$, this defines a
transversal measure defined by a $(n-p)$-form $\eta= i_{{\bf X}_p} (\mu)$. The de Rham current is
$$
C_{\mu, {\bf X}_p} (\beta) = \sum_i \int_{{\bf D}^{n-p}_i} \eta
\int_{{\bf D}_i^p \times \{y\}} \beta
$$
\begin{equation}
= \int_M i_{{\bf X}_p} (\mu) \wedge
\beta.
\label{current}
\end{equation}
This determines a closed current when $\mu$ is invariant under
the flow described by ${\bf X}_p$, here $\beta$ is a $p$-form.

Now we will give some useful results concerning the existence of higher
dimensional asymptotic cycles (for more details see \cite{SH}).
Consider the action of a connected Lie group $L$ on a smooth
compact oriented manifold $M$, whose orbits are tangent to the
orbits of dimension $p$.  A \textit{quantifier} is a continuous
field of $p$-vectors on $M$ everywhere tangent to the orbits and
invariant under the action of $L$. A quantifier is said to be
\textit{positive} if it is nowhere vanishing and determines the
orientation of the tangent space. A \textit{preferred action} is
an oriented action of a connected Lie group $L$ such that for any
$x \in M$ the isotropy group $D_x$ of $x$ is a normal subgroup of
$L$ and $L/D_x$ is unimodular.

It was proved in \cite{SH} that a preferred action possesses a
positive quantifier. Given a positive quantifier we define a
$1-1$ correspondence between a finite invariant measure $\mu$ and
a transversal invariant measure. An important result which will be used in
the next sections is a theorem that states as follows. If ${\bf X}_p$ is a positive definite
quantifier and $\mu$ is an invariant measure described by a $n$-form, then $i_{{\bf X}_p}
(\mu)$ is a closed $(n-p)$-form and the asymptotic cycle $A_\mu$ can be obtained by Poincar\'e
duality of an element of $H^{n-p} (M, \mathbb{R})$ determined by $i_{{\bf X}_p} (\mu)$.

If $A_\mu$ is an asymptotic cycle i.e. is an element of $H_p (M,
\mathbb{R})$, the theorem give us an explicit form to construct
asymptotic cycles if the above conditions are satisfied. This is
not the only way to specify a foliation, in \cite{S} Sullivan
defines structures of $p$-cones and operators acting over
vectors on these cones.

\subsection{Asymptotic Hopf Invariant for a Flow}

Using the idea of asymptotic cycle we define asymptotic average
linking number as in Ref. \cite{Arnold}. Let $M$ be a
closed and orientable $3$-manifold with volume form (invariant
measure) $\mu$, which we will assume that is normalized as $\int_M \mu =
1$. Consider a vector field $X$ that induces a flow $f_t$, which
satisfies $\mathcal{L}_X \mu = 0$ (divergence-free vector field) which is homologically
trivial i.e., there exist a $1$-form $\alpha$ such that $d\alpha = i_X
(\mu)$. For a closed $1$-form the winding cycle is zero for a
homologically trivial vector field
$$
\Psi_\mu = \int_M \omega (X) \mu
= \int_M \omega \wedge i_X(\mu)
$$
\begin{equation}
\quad = \int_M d(\omega \wedge \alpha) =0.
\end{equation}
Under this situation the asymptotic linking number exists and is
well defined \cite{Arnold,Arnold-book,vogel}

\begin{equation}
\label{aln}
L (\Gamma_{x_1}, \Gamma_{x_2}) = \lim_{T_1, T_2 \to
\infty} \frac{L(\widetilde\Gamma_{T_1, x_1},
\widetilde\Gamma_{T_2, x_2})}{T_1 T_2},
\end{equation}
where $\widetilde\Gamma_{t, x_i} = [x_i, f_t (x_i)] \cup
\gamma_{f_t(x_i), x_i}$, with $[x_i, f_t (x_i)]$ being the
oriented arc of trajectory from $x_i$ to $f_t(x_i)$ and $\{
\gamma_{f_t(x_i), x_i} \}$ is a set of regular curves that join
$f_t(x_i)$ and $x_i$ with $i=1,2$.\\

The mean value of the asymptotic linking number of a pair of
trajectories (average Hopf invariant) of a homologically trivial
vector field $X$ on $M$ is given by
\begin{equation}
\label{aaln}
L(X) = \int_{M \times M}  \! \! \! \! L
(\Gamma_{x_1}, \Gamma_{x_2}) \; d v (x_1) d v (x_2).
\end{equation}
Finally Arnold proved \cite{Arnold,Arnold-book} that this number is equal to
\begin{equation}
\label{aaln1}
L(X) = \int_M \alpha \wedge d \alpha,
\end{equation}
where $d \alpha = i_X (\mu)$.  The explicit form for  $L(X)$ is given by

\begin{equation}
\label{aaln2}
 L(X) = \int_{M \times M} \! \! \! \! \frac{
\varepsilon_{ijk}X^i(x_1) X^j(x_2) (x_1-x_2)^k}{| x_1 - x^2|^3} \,
d v (x_1) d v (x_2).
\end{equation}

\section{Asymptotic Linking Numbers in Higher Dimensions}

In this section we generalize some of the material revisited in Sec.
2 to higher dimensions.  In order to do that we will use the definition of asymptotic cycles
in higher dimensions from \cite{SH}. Our main motivation is the extension of the ideas of dynamical systems to string theory.

\subsection{Average Asymptotic Linking Number}

Now we proceed to give the definition of the asymptotic linking number in higher dimensions
using the standard $BF$ action without cosmological constant term (\ref{action1}) given in section 2.

Let us consider $M$ to be a $n$-dimensional manifold such that $H_p (M, \mathbb{R}) = 0$. $M$ is also equipped with an invariant flow volume form $\mu$ and a divergence-free $p$-vector field ${\bf X}_p=X_1 \wedge \cdots \wedge X_p$  i.e. $\mathcal{L}_{{\bf X}_p} \mu =0$ (see \cite{Holm,vaisman} for notation, conventions and properties regarding multi-vector fields). Here each $X_i$ with $i=1, \cdots ,p$ is a vector field ($1$-vector field). For a $p$-vector field the Lie derivative is defined as: $\mathcal{L}_{{\bf X}_p} = [i_{{\bf X}_p}, d],$ where $i_{{\bf X}_p}$ is defined by $i_{{\bf X}_p}= i_{X_1 \wedge \cdots \wedge X_p} = i_{X_1} \circ \cdots \circ i_{X_p}$
is the contraction, notice that all $i_{X_i}$'s commutes among themselves). If $N= \partial W$ is a null homologous $(p'=n-p-1)$ dimensional submanifold of $M$, $N$ is boundary of some $(n-p)$-dimensional manifold.

Motivated by \cite{SA,SH,RS,VV} we define the following asymptotic observable $\widetilde{O}_{{\bf X}_p}$ as a current
\cite{dRham}
\begin{equation}
\label{aobvs}
\widetilde{O}_{{\bf X}_p} = \int_M i_{{\bf X}_p} (B)  \mu_{{\bf X}_p} (p_1).
\end{equation}
Then the two-point correlation function is consequently

\begin{eqnarray}
\left\langle \widetilde{O}_{{\bf X}_p}(p_1)  \cdot O_N (p_2)
\right\rangle = \left\langle \int_M i_{{\bf X}_p}(B) d\mu(p_1) \cdot \int_N C(p_2) \right\rangle \quad \quad \quad \quad \quad \quad \quad \quad \quad \quad  \nonumber \\
= \frac{\int {\cal D} B {\cal D} C \  \exp \left[ i\int_M B \wedge
d C \right] \int_M i_{{\bf X}_p}(B)  \mu(p_1)   \int_N C
(p_2)}{\int {\cal D} B {\cal D} C \ \exp \left[ i\int_M B \wedge d C \right] }.
\label{ultima1}
\end{eqnarray}
These observables are invariant under the gauge transformations (\ref{gauge}). In the present case it can be written as
\begin{equation}
\int_M i_{{\bf X}_p} (B') \mu (p_1) = \int_M i_{{\bf X}_p} (B)  \mu_{{\bf X}_p} (p_1) + \int_M i_{{\bf X}_p} (dv) \mu_{{\bf X}_p} (p_1),
\end{equation}
where $v$ is a $(p-1)$-form. The second term vanishes due the theorem 2A in  \cite{SH}, which states that the current is closed if and only if $\mu$ is an invariant measure. The second observable $\int_N C$  is clearly gauge invariant.

In order to compute the rhs expression of (\ref{ultima1}), as in the case without flows \cite{Horowitz:1989km}, we again split the forms using the Hodge decomposition theorem.  If $B$ is a $p$-form we can
write down it uniquely as: $B = B^0 + d\phi + \delta \xi$, where $B^0$ is the harmonic part,  $d\phi$ and $\delta \xi$ are the longitudinal and transversal parts of $B$ respectively, $d$ is the usual exterior derivative and $\delta=(-1)^{pn+n+1} *d*$ the adjoint. Here $\phi$ is a $(p-1)$-form and $\xi$ is a $(p+1)$-form, $*$ is the Hodge operator on $M$ such that $\ast^2= (-1)^{p(n-p)}$.

To evaluate this integral, we split the measure in the following
way with the aid of the Hodge theorem: $\mathcal{D} B = \mathcal{D} B^T \mathcal{D} B^L \mathcal{D} B^0$ and similarly for ${\cal D}C$. Only the transverse part of the $BF$ action and observables contributes, i.e.
$\widetilde{O}_{{\bf X}_p} = \int_M i_{{\bf X}_p} (B^T) \mu$ and $O_N = \int_N C^T$ therefore we integrate out all longitudinal and harmonic modes leaving only the transverse ones $\mathcal{D} B^T \mathcal{D} C^T.$

Consider $\lambda^2 \not= 0$ to be an eigenvalue of the laplacian $\Delta_p = \delta  d + d \delta$ let $\Lambda_p$ the space of $p$ eigenforms ($\Delta B = \lambda^2 B$). This space is partitioned in the transverse $\Lambda^T_p$ and longitudinal $\Lambda^L_p$ parts. Following to \cite{Horowitz:1989km} we have an isomorphism between $\Lambda_p$ and $\Lambda_{n-p}$ due to the Hodge dual maps $p$-forms with $\lambda^2$ to $(n-p)$-forms with the same eigenvalue. Also we have that the codifferential is a mapping from $\Lambda_p^L$ to $ \Lambda_{p-1}^T$ and the differential $d$ maps from $\Lambda_p^T$ to $\Lambda_{p+1}^L$. Thus we can define the following map $\lambda^{-1} \ast d : \Lambda_p^T \to \Lambda_{n-p-1}^T$, which define an isomorphism.

Let $\{B_j\}$ and $\{C_j\}$ be  basis of normalized transverse eigenforms which satisfies $\langle B_j \vert B_k \rangle = \int_M B_j \wedge \ast B_k = \delta_{jk}$. If one takes the following choice for $C_j = (-1)^{n-p-1} \lambda_j^{-1} \ast d B_j$, then the $C_j$ are also orthonormal. Let us take $B$ and $C$ in their respective expansion of the basis $B=\sum_j b_j B_j$ and $C= \sum_j c_j C_j$, then the action can be written as: $S_{BF} = \langle B \vert C \rangle = \sum_j \lambda_j b_j c_j$ due to the normalization of $B_j$ and $C_j$.  The transverse measure of the path integral takes the following form $\mathcal{D} B^T \mathcal{D} C^T = \prod_j db_j
\prod_k dc_k$. Substituting and integrating (\ref{ultima1}) this yields
$$
H({{\bf X}_p},N) := {1\over i}\left\langle \widetilde{O}_{{\bf X}_p}(p_1)  \cdot O_N (p_2)
\right\rangle
$$
\begin{equation}
\label{alink}
= \sum_j \lambda_j^{-1}\int_M i_{{\bf X}_p} (B_j) \vert_{p_1} \mu (p_1) \cdot \int_N C_j (p_2).
\end{equation}
Using the properties of $i_{X_1 \wedge \cdots \wedge X_p}$ it is easy to see that

\begin{equation}
\label{derivation}
i_{X_1 \wedge \cdots \wedge X_p} B_j \wedge \mu - (-1)^{{p \over 2}(3+p)} B_j \wedge i_{X_p \wedge \cdots \wedge X_1}\mu=0.
\end{equation}
Substituting the last equality in $(\ref{alink})$, we obtain

\begin{equation}
H({\bf X}_p,N) =  (-1)^{{p \over 2}(3+p)}\sum_j \lambda_j^{-1} \int_M  B_j \vert_{p_1} \wedge i_{{\bf X}} \mu (p_1) \cdot \int_N C_j (p_2).
\end{equation}
If ${\bf X}_p$ is a divergence-free $p$-vector field, this implies that $\eta = i_{{\bf X}_p} \mu$ is closed. Since $H_p(M,\mathbb{R})$ is trivial then by Poincar\'e duality there exists $\alpha \in \Omega^{n-p-1} (M) $ such that $\eta = d \alpha'$, where $\alpha' := (-1)^{{p \over 2}(3+p)} \alpha$. Then $N$ is null-homologous i.e. $N= \partial W$, where $W$ is a $(n-p)$-manifold we get

\begin{eqnarray}
H({\bf X}_p,N) &=& \sum_j \lambda_j^{-1} \int_M  B_j \vert_{p_1} \wedge \eta (p_1) \cdot \int_N C_j (p_2) \nonumber \\ &=& \sum_j \lambda_j^{-1} \int_M B_j \vert_{p_1} \wedge \eta (p_1) \cdot \int_W d C_j (p_2).
\end{eqnarray}
Using $dC_j = \lambda_j \ast B_j$  we obtain the following expression

\begin{equation}
H ({\bf X}_p,N) =  \sum_j \int_M \int_W B_j (p_1) \ast  B_j (p_2) \wedge \eta (p_1).
\end{equation}
Using the completeness relation for the eigenforms i.e.

$$
\sum_j B^T_j(p_1) \ast B^T_j(p_2) + \sum_j B^L_j(p_1) \ast B^L_j(p_2) + \sum_j B^0_j(p_1) \ast B^0_j(p_2) = \delta (p_1, p_2) v(p_1) \cdot v(p_2),
$$
where $v(p_1)$ and $v (p_2)$ are $p$ and $n-p$ ``{\it volume}" forms in a $p$ foliation of $M$. Since the longitudinal and harmonic forms do not contribute to the path integral we can include them and integrate over $M$, we have the following result
$$
H ({\bf X}_p,N) =  \int_M \int_W  \delta (p_1, p_2) v(p_1) \cdot v(p_2) \wedge \eta (p_1)
$$
$$
 = \int_W \eta (p_1)
$$
\begin{equation}
\label{thopf1}
= \int_N \alpha'.
\end{equation}
This is precisely a Hopf type integral. It is immediate to note that for the case of $p=1$ we get the result $\int_N \alpha$ by  Kotschick and Vogel \cite{KV}. For this case $H_1 (M, \mathbb{R})$ is trivial, $X$ is a divergence-free vector field, $\alpha$ is a $(n-2)$-form, which satisfies $i_X (\mu) = d \alpha$.


\subsection{Asymptotic Intersection of Two Flows}
In this subsection we will calculate a correlation function of two asymptotic observables. This leaves us to define  the asymptotic linking number between two flows
of dimension $p$ and $p'$. After that we will calculate the asymptotic self-intersection and finally extract the Hopf invariant in higher dimensions.

Let $\widetilde{O}_{{\bf X}_p}$ and $\widetilde{O}_{{\bf Y}_{p'}}$  be two asymptotic observables defined as in Eq. (\ref{aobvs}). We will consider that each observable has it own flow invariant measure
$\mu_{{\bf X}_p}$ and $\mu_{{\bf Y}_{p'}}$, where ${\bf X}_p=X_1 \wedge \ldots \wedge X_p$ and ${\bf Y}_{p'}= Y_1 \wedge \ldots \wedge Y_{p'}$. Suppose $M$ has trivial $p$-th and $(p+1)$-th homology groups then the two point correlation function to calculate is the following:

\begin{eqnarray}
\left\langle \widetilde{O}_{{\bf X}_p}(p_1)  \cdot \widetilde{O}_{{\bf Y}_{p'}} (p_2)
\right\rangle = \frac{\int    \mathcal{D} B \mathcal{D}
C \  \exp \left[ i\int_M B \wedge d C \right] \widetilde{O}_{{\bf X}_p} (p_1)  \cdot \widetilde{O}_{{\bf Y}_{p'}} (p_2) }{\int  \mathcal{D} B \mathcal{D} C
\  \exp \left[ i\int_M B \wedge d C \right] } \nonumber \\
= \frac{\int {\cal D} B {\cal D} C \  \exp \left[ i\int_M B \wedge
d C \right] \int_M i_{{\bf X}_p}(B)  \mu_{{\bf X}_p}(p_1) \cdot \int_M i_{{\bf Y}_{p'}}(C)
\mu_{{\bf Y}_{p'}}(p_2)}{\int {\cal D} B {\cal D} C \ \exp \left[ i\int_M B
\wedge d C \right] }. \label{slink}
\end{eqnarray}
Remember that to the path integral only contributes the transversal part since the longitudinal modes and the harmonic decouples from the transverse ones and can be directly integrated out (\ref{slink}) and it reduces to

$$
\left\langle \widetilde{O}_{{\bf X}_p}(p_1)  \cdot \widetilde{O}_{{\bf Y}_{p'}} (p_2)
\right\rangle
$$
\begin{eqnarray}
=  i \sum_j \lambda_j^{-1} \int_M i_{{\bf X}_p}(B) \vert_{p_1} \mu_{{\bf X}_p}(p_1) \cdot
\int_M i_{{\bf Y}_{p'}} (C) \vert_{p_2} \mu_{{\bf Y}_{p'}} (p_2).
\end{eqnarray}
Dividing by $i$ we define the asymptotic linking number $H({\bf X}_p ,{\bf Y}_{p'})$ of ${\bf X}_p$ and ${\bf Y}_{p'}$  as follows
$$
H ({\bf X}_p,{\bf Y}_{p'}) := {1 \over i}\left\langle \widetilde{O}_{{\bf X}_p}(p_1)  \cdot
\widetilde{O}_{{\bf Y}_{p'}} (p_2)
\right\rangle
$$
\begin{eqnarray}
=\sum_j \lambda_j^{-1} \int_M B_j \vert_{p_1} \wedge \eta_1 (p_1) \int_M C_j \vert_{p_2} \wedge \eta_2 (p_2),
\label{intersection}
\end{eqnarray}
where we apply (\ref{derivation}) to the two integrals. We also use the facts that the fields are divergence-free to prove that  $\eta_1 = i_{X_1 \wedge \cdots \wedge X_p}(\mu_{{\bf X}_p})$ and $\eta_2 = i_{Y_1 \wedge \cdots \wedge Y_{p'}} (\mu_{{\bf Y}_{p'}})$ are closed
and providing our assumption that $H_p (M ,\mathbb{R})= H_{p+1} (M, \mathbb{R})=0$ they are exact i.e., $\eta_1 = d \alpha'_1$ and $\eta_2 = d \alpha'_2$, where $\alpha'_1= (-1)^{{p \over 2}(3+p)} \alpha_1$ and $\alpha'_2= (-1)^{{p' \over 2}(3+p')} \alpha_2$ with $\alpha_1 \in \Lambda^{p}(M)$ and $\alpha_2 \in \Lambda^{p'}(M)$.

Integrating by parts the second integral, using again the fact that $dC_j = \lambda_j  \ast B_j$, and summing over all the states (including the longitudinal and harmonic forms) we get
$$
H ({\bf X}_p,{\bf Y}_{p'}) =\int_M \int_M \delta (p_1, p_2) \wedge \eta_1 (p_1)  \alpha_2 (p_2)
$$
\begin{equation}
\label{tipohopf}
=\int_M  d\alpha'_1 \wedge \alpha'_2.
\end{equation}
Thus we can think of this expression as a Hopf like invariant or as the asymptotic intersection associated to the vector fields ${\bf X}_p$ and ${\bf Y}_{p'}$. Let us see some considerations.

As an example we would like to compute the asymptotic self-linking number of a divergence-free $p$-vector field ${\bf X}_p= X_1 \wedge \cdots \wedge X_p$ i.e. we want $H({\bf X}_p,{\bf X}_p)$ with ${\bf X}_p={\bf Y}_{p'=p}$. If $M$ is of dimension $n=2p+1$ where for $p \geq 1$ by dualities between homology and cohomology we relax the condition only to assume $p$-th homology group will be zero. From (\ref{tipohopf}) then the expectation value  is:

\begin{equation}
H({\bf X}_p,{\bf X}_p) = \int_M  \alpha \wedge d \alpha,
\label{hopf-helicity}
\end{equation}
where $\alpha$ is a $(p-1)$-form given by the equation $i_{{\bf X}_p} ( \mu ) = d \alpha$. This expression is exactly the Hopf invariant (or helicity) in dimension greater than three  \cite{bott-tu} or the higher dimensional helicity of a $p$-vector field ${\bf X}_p$ (it can be associated to a $p$-foliation since the observables defines $p$-currents on $M$) equipped with an invariant measure $\mu$, associated to a $p$-foliation of $M$. The involved $p$-vector field ${\bf X}_p$ represents a choice of a framing on the tangent bundle of $M_n$ and consequently (\ref{hopf-helicity}) does not have the coincidence singularity and it is perfectly regular. 

Clearly for the case when we take $n=3$ and $p=1$, we have the classical result by Arnold, the asymptotic Hopf invariant (or helicity) of a incompressible flow on a $3$-manifold \cite{Arnold}.

\subsection{Generalized Linking Numbers for Flows}

Now we want to calculate the asymptotic generalized linking number of a $p$-vector field. In order to do this we recall the asymptotic observable defined as in Eq. ($\ref{aobvs}$).

\vskip 1truecm
\noindent
{\it One Flow}

Now we consider the intersection of a flow ${\bf X}_p$ with a null homologous submanifold $N= \partial W$ in $M$ of dimension $p'$. Similarly we assume $H_p (M, E)$ is trivial where $E$ is a trivial gauge bundle of with  $SU(N)$ structure group. Now we define the generalized asymptotic linking number as the following correlation function:

\begin{equation}
H_A({\bf X}_p,N) =  \frac{1}{i} \left\langle \int_M i_{{\bf X}_p}(\mathcal{B}) \mu(p_1) \cdot  \int_N {\rm P} \exp \left( \int_{p_1}^{p_2} A \right) \mathcal{C}(p_2) \right\rangle,
 \end{equation}
where $\mathcal{B}\in \fund$ and $\mathcal{C} \in \antifund$, $A \in {\bf adj}=(\fund,\antifund)$. The observable will be gauge invariant and the correlation function will be a scalar. The expectation value is invariant since the currents are invariant under the gauge transformations, provided that $\gamma_0$ is a curve that join $p_1$ and $p_2$. In this case also the transverse modes decouples from the longitudinal and the transverse ones and they factorizes in the path integral in such a way that they can be integrated out. We have
\begin{equation}
H_A ({\bf X}_p, N) = \sum_j \lambda_j^{-1}\int_M i_{{\bf X}_p} (\mathcal{B}_j) \vert_{p_1} \mu (p_1) \cdot  \int_N
{\rm P} \exp \left( \int_{p_1}^{p_2} A \right) \mathcal{C}_j (p_2)
\end{equation}
or
\begin{equation}
H_A ({\bf X}_p, N) = \sum_j \lambda_j^{-1}\int_M \mathcal{B}_j \vert_{p_1}  \wedge \eta (p_1) \cdot  \int_N {\rm P} \exp \left( \int_{p_1}^{p_2} A \right) \mathcal{C}_j (p_2).
\end{equation}
Since $N$ is null homologous using Stokes theorem and taking $D \mathcal{C}_j = \lambda_j \ast \mathcal{B}_j $, we obtain

\begin{equation}
H_A ({\bf X}_p, N) = \sum_j  \int_M \mathcal{B}_j \vert_{p_1}  \wedge \eta (p_1) \cdot \int_W  {\rm P} \exp \left( \int_{p_1}^{p_2} A \right) \ast \mathcal{B}_j (p_2)
\end{equation}
in the last equation we have use the fact that $A$ is flat (i.e. $F_A=dA +A\wedge A=0$). In order to proceed the computation we incorporate the spurious modes, this yields

\begin{equation}
H_A ({\bf X}_p, N) =  \int_M \int_W \delta(p_1,p_2) v(p_1) \cdot v(p_2) \cdot  {\rm Tr \ P}  \exp \left( \int_{p_1}^{p_2} A \right) \eta (p_1).
\end{equation}
Integrating over $p_1$ this equation reduces to

\begin{equation}
H_A ({\bf X}_p, N) =   \int_W  {\rm Tr \ P}  \exp \left( \oint_{\gamma} A \right) \eta,
\label{conn1}
\end{equation}
where $\gamma$ is build as follows: take a curve from $p_2$ in $W$ to a point $u$ in the Poincar\'e dual of $\eta$. Then use the curve $\gamma_0$ from $u$ to $v$ in $V$ and finally take a curve from $u$ to $p_2$ that is contained in $W$.

\vskip 1truecm
\noindent
{\it Two Flows}

Now consider two divergence-free vector fields and $H_{p+1}(M,\IR)=H_{p'+1}(M,\IR)=0$. The first one a $p$-vector field ${\bf X}_p$ and $p'$-vector field ${\bf Y}_{p'}$. Every field has his own flow invariant measure, let say $\mu_{{\bf X}_p}$ and $\mu_{{\bf Y}_{p'}}$ respectively. We extend (\ref{slink}) and then the generalized asymptotic linking number $H_A ({\bf X}_p, {\bf Y}_{p'})$ as follows

\begin{equation}
H_A ({\bf X}_p, {\bf Y}_{p'}) = \frac{1}{i} \left\langle \int_M i_{{\bf X}_p} (\mathcal{B}) \mu_{{\bf X}_p} (p_1)  \cdot \int_M {\rm P} \exp \left( \int_{p_1}^{p_2} A \right) i_{{\bf Y}_{p'}} (\mathcal{C}) \mu_{{\bf Y}_{p'}} (p_2) \right\rangle.
\end{equation}
Following a similar procedure as in the previous cases we get

\begin{equation}
H_A ({\bf X}_p,{\bf Y}_{p'})  =  \sum_j \lambda_j^{-1} \int_M i_{{\bf X}_p} (\mathcal{B}_j) \mu_{{\bf X}_p} (p_1) \cdot  \int_M  {\rm P} \exp \left ( \int_{p_1}^{p_2} A   \right)   i_{{\bf Y}_{p'}} (\mathcal{C}_j) \mu_{{\bf Y}_{p'}} (p_2) .
\end{equation}
Using the same identities it yields

\begin{equation}
H_A ({\bf X}_p,{\bf Y}_{p'})  =  \sum_j \lambda_j^{-1} \int_M  \mathcal{B}_j \vert_{p_1} \wedge  \eta_{{\bf X}_p} (p_1) \cdot \int_M  {\rm P} \exp \left ( \int_{p_1}^{p_2} A   \right)  \mathcal{C}_j \vert_{p_2} \wedge \eta_{{\bf Y}_{p'}} (p_2).
\end{equation}
Using again the relation: $D_A {\cal C}_j =\lambda_j * {\cal B}_j$ and the completeness relation and integrating out  rhs is  rewritten as

\begin{equation}
H_A ({\bf X}_p, {\bf Y}_{p'}) =  \int_M  {\rm Tr \ P} \exp \left ( \oint_\gamma A   \right) d \alpha_{{\bf X}_p} \wedge \alpha_{{\bf Y}_{p'}}.
\end{equation}
Take $\eta_{{\bf X}_p}$ and $\eta_{{\bf Y}_{p'}}$, by Poincar\'e duality there are dual homology cycles $\Gamma_{{\bf X}_p}$ and $\Gamma_{{\bf Y}_{p'}}$ respectively which are trivial homology classes, therefore $\gamma$ is a curve starting in $p_2 \in W$ whose boundary is $\Gamma_{{\bf X}_p}$ to a point $u \in \Gamma_{{\bf Y}_{p'}}$ completely contained in $\Gamma_{{\bf Y}_{p'}}$. Then take a curve $\gamma_0$ from $u$ to $v$ in $\Gamma_{{\bf X}_p}$ and then go back to the point $p_2$ from $v$ contained completely in $W$.

Now consider de self-intersection i.e., take ${\bf X}_p = {\bf Y}_{p'=p}$ to be a $p$-vector field and $H_p (M, \mathbb{R})=0$. The dimension of $M$, $n$ is $2p + 1$ then the expression reduces to

\begin{equation}
H_A ({\bf X}_p, {\bf X}_p) =  \int_M  {\rm Tr \ P} \exp \left ( \oint_\gamma A   \right) \alpha  \wedge d\alpha .
\label{conn2}
\end{equation}
We can think in this expression as the generalized asymptotic Hopf invariant associated to the field ${\bf X}_p$.

\section{Higher Dimensional Asymptotic Jones-Witten Invariants}
We will give a briefly overview of the asymptotic Jones-Witten invariants \cite{VV}, and give an extension to higher dimensions.

\subsection{Jones-Witten Invariants for Flows on 3-manifolds}

Let us consider a closed 3-manifold $M$ with a divergence-free vector field $X$ and an
invariant probability measure $\mu$. Let $A$ be a connection
on the $G$-principal bundle: $P \stackrel{\pi}{\to} M$. To define the asymptotic
Jones-Witten invariant basically we will modify the the Wilson line and use the extended version, the {\it asymptotic Wilson line}. We will interpret this in terms of asymptotic homology cycles.

For the case of an abelian gauge group let us take $G=U(1)$, the asymptotic holonomy is the
limit of the Wilson loop of $Hol_{\widetilde \Gamma_{t,p}} (A) = P
\exp \int_{\Gamma_{t,p}} A$, i.e. the asymptotic holonomy is
defined as ${\rm Hol}_{\Gamma_p} (A) = \lim_{t \to \infty} \exp
\int_{\tilde\Gamma{t,p}} \frac{1}{t}A$, next we will define the
{\it average holonomy} of a $U(1)$-connection, over all the asymptotic cycles as
$$
{\rm Hol}_{X,\mu}(A)= \exp \int_M A(X) \mu
$$
\begin{equation}
=\lim_{t
\to \infty} \exp \int_M \bigg(\int_{\widetilde{\Gamma}_{t,p}}
\big({1\over t} A \big)\bigg)  \mu(p).
\end{equation}
For the non-abelian case, for instance, for $G=SU(2)$ Verjovsky and Vila Freyer redefine the asymptotic Wilson line in terms of the monodromy (see \cite{VV, Arnold-book} for details). In the present paper we will limited ourselves to the abelian case, thus we going come back to this case. In this case the asymptotic Jones-Witten invariants are defined by the following functional

\begin{equation}
W_{X,\mu}(k) = \int_{{\cal A}/{\cal G}} {\cal D}A \ \exp \bigg({ik
\over4 \pi} \int_M A \wedge dA \bigg) \cdot {\rm
Hol}_{X,\mu}(A), \label{jwinv}
\end{equation}
where $k\in \IZ$,  ${\cal A}$ is the space of all $U(1)$-flat connections on $P$ and
${\cal G}$ the gauge group. The computation of $W_{X,\mu}(k)$ leads to \cite{VV, Arnold-book}:

\begin{equation}
W_{X,\mu}(k) = c(M) \exp \bigg\{{2 \pi i \over k} \int_{M}
\alpha \wedge d \alpha \bigg\}
\end{equation}
where $c(M)$ contains the Ray-Singer torsion and the term of the exponential is precisely the asymptotic Hopf invariant, remember $i_X (\mu) = d \alpha$.

\subsection{Jones-Witten Invariants for a High Dimensional Flow}
In analogy to the previous section we define a Jones-Witten invariant for a flow in a $n=2p+1$ dimensional manifold, with $H_p (M, \mathbb{R} )= 0$. Take a divergence-free $p$-vector field ${\bf X}_p$ and consider the $BF$ theory taking a $p$-form $B=C$ then the action is the Chern-Simons functional $S_{BF}= \int_M B\wedge dB$. We define the following asymptotic observable

\begin{equation}
\label{expobs}
\mathcal{O}_{{\bf X}_p,\mu}  = \exp \left( i \int_M i_{{\bf X}_p} (B) \mu \right).
\end{equation}
Then the asymptotic Jones-Witten invariants in $n=2p+1$ dimensions are given by
$$
{\cal W}_{{\bf X}_p}(\mu) := \left\langle \mathcal{O}_{{\bf X}_p, \mu} \right\rangle
$$
\begin{equation}
= {\cal N} \int  {\cal D}B \ \exp \bigg( i \int_M B \wedge dB \bigg) \cdot \mathcal {O}_{{\bf X}_p,\mu},
\end{equation}
where ${\cal N}$ is the normalization factor.

As in previous cases we use the Hodge decomposition theorem and again only the transverse modes will contributes to the path integral. Recalling that ${\bf X}_p$ is a divergence-free vector field we have $i_{{\bf X}_p} (\mu) = d \alpha$. Then we have

\begin{equation}
 {\cal W}_{{\bf X}_p}(\mu) = {\cal N} \int  {\cal D}B^T \ \exp \bigg(  i\int_M B^T  \wedge dB^T + B^T \wedge d \alpha ',  \bigg)
\end{equation}
where $\alpha ' = (-1)^{{p \over 2}(3+p)} \alpha$.

The reason is the same as the previous case where the longitudinal and harmonic modes factor out and are absorbed in the normalization factor. Because the current (\ref{expobs}) is closed in the de Rham sense (invariant under gauge transformation as in section $4$) we take a basis to expand the transverse forms which satisfies $\ast d B_j = \lambda_j B_j$, where  $\lambda_j^2$ are the eigenvalue of the laplacian. Expanding $B^T = \sum_j b_j B_j$, with $b_j$'s being scalars and $\langle B_i \vert B_k \rangle = \int_M B_i \wedge \ast B_k = \delta_{ik}$. After some computations we get

\begin{equation}
 {\cal W}_{{\bf X}_p}(\mu)= {\cal N} \int \prod_j db_j  \exp \bigg( i  \sum_j \lambda_j ( b_j^2 + \alpha'_j b_j )   \bigg),
\end{equation}
where $\alpha_j= \langle \alpha \vert B_j \rangle$. Finally we obtain

\begin{equation}
 {\cal W}_{{\bf X}_p}(\mu) = C(M)  \exp \bigg(- \frac{i}{4}  \sum_j \lambda_j  \alpha'^2_j    \bigg),
\end{equation}
where $C(M)$ is a constant that can be removed by normalizing the expectation value. We observe that $\lambda_j \langle \alpha ' \vert B_j \rangle = \int_M d\alpha ' \wedge B_j$, where we did the integration, then the argument of the exponential is $\sum_j \int_M d \alpha ' \wedge B_j \int_M \alpha ' \wedge \ast B_j = \int_M \alpha \wedge d \alpha$, where we use the completeness relation and finally we get the expression
$$
{\cal W}_{{\bf X}_p}(\mu)= C(M)   \exp \bigg(- \frac{i}{4} \int_M \alpha \wedge d \alpha  \bigg)
$$
\begin{equation}
= C(M)   \exp \bigg(- \frac{i}{4} H ({\bf X}_p)  \bigg),
\label{highJW}
\end{equation}
where $H({\bf X}_p)=H ({\bf X}_p, {\bf X}_p)$ is the high dimensional asymptotic Hopf invariant (\ref{hopf-helicity}).
The expression (\ref{highJW}) is a topological invariant, i.e., the Jones-Witten invariants in $n=2p+1$ dimensions. They only depend on the flow ${\bf X}_p$ (or the foliation of the manifold M, associated to the flow generated by ${\bf X}_p$) and the invariant measure $\mu$.

\section{Asymptotic Invariants in String Theory}

It is known that $BF$ theories arises in natural way in supergravity in eleven dimensions. In addition to the Einstein-Hilbert action in eleven dimensions we have:
\begin{equation}
S_{sugra}=\cdots + \int_{M_{11}} \big(  G_4 \wedge * G_4 + C_3 \wedge G_4 \wedge G_4 \big) + \ {\rm anomalous \ terms}
\end{equation}
where $G_4= dC_3$. This is a theory with M2 branes $W$ coupled to the three-form $C_3$. The observables are the form $\int_W C_3$ and one can compute, for instance, the two-point correlation functions $ \left \langle \int_W C_3 (x) \cdot \int_W C_3 (y)   \right \rangle$

In Type II  superstring theory we have the Chern-Simons coupling which is the coupling between the RR-fields $C_p$ and the other fields of the theory. In addition to the Dirac-Born-Infeld action we will have the CS action given by:
$$
I_{CS}= \sum_p C_p \wedge \exp \bigg(2 \pi  \alpha '(B+ F)\bigg)
$$
\begin{equation}
= \int_{W_p }C_p + \int_{W_p} B \wedge C_{p-2} + \int_{W_p} F \wedge C_{p-2} + \cdots
\label{cs-action}
\end{equation}
The action would be any of the two kinds of Type II theories \cite{Polchinski:1998rr} $S_{II}$ contains a term of the CS form \begin{equation}
S_{IIA}= \cdots + \int_{M_{10}} B \wedge F_4 \wedge F_4 + \cdots,
\end{equation}
where $F_4= dC_3$ is the field strength of $C_3$ or
\begin{equation}
S_{IIB} = \cdots + \int_{M_{10}} C_4 \wedge H_3 \wedge F_3 + \cdots,
\end{equation}
where $H_3=dB$. In Type IIA $p=1,3,5,7,9$ and in Type IIB $p=0,2,4,6,8$. Thus we can define the correlation function corresponding to two non-intersecting D-branes of world-volumes $W$ and $W'$ with RR-fields $C_p$ and $C_{p'}$ respectively, thus we have
\begin{eqnarray}
 \left \langle \int_W C_p (x) \cdot \int_{W'} C_{p'} (y)   \right \rangle =
 \frac{\int  \mathcal{D}
C \  \exp \left[ iS_{IIA,B} \right] \int_W C_p (x) \cdot \int_{W'} C_{p'} (y)}
{\int \mathcal{D} C
\  \exp \left[ iS_{IIA,B} \right] }  \label{correla}
\end{eqnarray}
with $x \in W$ and $y \in W'$. Of course the suitability of them depend on the possibility to solve the integration on the rhs. However in the present section we will not follow this path and we only study the consistency of the lhs and the definition of appropriated asymptotic observables. The observables $\int_W C_p$ are gauge invariant under gauge transformations $C_p \to C_p + d\Lambda_{p-1}$, they depend only of the homology class $[W]$ of $M$.

Now let us consider dynamical D-branes. It is known that one can scatter open or closed strings by  D-branes but the D-branes as a dynamical objects also can be scattered by themselves. In the strong coupling limit $g_S \to \infty$, the D-branes are light objects and can be scattered by a center of forces \cite{McAllister:2004gd}. In this section we would like to study this system from the point of view of dynamical systems. To be more precise we describe the motion of a D-brane on a foliation $M_{1,9}= \IR \times {\bf D}^p \times {\bf D}^{9-p}$, instead of a manifold $M_{1,9}=\IR \times \IR^9$. This corresponds physically to have a D$p$-brane moving in the $M_{1,9}$ manifold along the transverse $(9-p)$ dimensions. In this case we can also define an asymptotic observable \cite{SA,SH,RS,VV} ${\cal O}_{\bf X}$ as:

\begin{equation}
\label{aobvsstring}
{\cal O}_{{\bf X}_p} = \int_M i_{{\bf X}_p} (C_p)  \mu_{{\bf X}_p}.
\end{equation}
Then the two-point correlation function is consequently
$$
H({\bf X}_p, {\bf Y}_{p'}):= {1 \over i}\left \langle  {\cal O}_{{\bf X}_p}(x) \cdot {\cal O}_{{\bf Y}_{p'}}(y)   \right \rangle
$$
\begin{equation}
= {1 \over i}\left \langle \int_M i_{{\bf X}_p} (C_p)  \mu_{{\bf X}_p}(x) \cdot  \int_M i_{{\bf Y}_{p'}} (C_{p'})  \mu_{{\bf Y}_{p'}} (y)\right \rangle.
\label{gauged}
\end{equation}
These quantities are invariant under the gauge transformations if the measures $\mu_{{\bf X}_p}$ and $\mu_{{\bf Y}_{p'}}$ are invariant under the $p$ and $p'$-flows respectively:
\begin{equation}
\int_M i_{{\bf X}_p} (C_p) \mu_{{\bf X}_p} (x) = \int_M i_{{\bf X}_p} (C_p)  \mu_{{\bf X}_p} (x) + \int_M i_{{\bf X}_p} (d\Lambda_{p-1}) \mu_{{\bf X}_{p}}  (y),
\end{equation}
and similarly for ${\bf Y}_{p'}$.

Again the second term vanishes by theorem 2A in  \cite{SH}. One would try to interpret Eq. (\ref{gauged}) as kind of linking number between the two flows ${\bf X}$ and ${\bf Y}$ associated to the propagation of the non-intersecting D$p$ and D$p'$ branes in the spacetime.

Now let us consider a couple of non-intersecting D-branes, a D$p$ and a D$p'$ of corresponding worldvolumes $W$ and $W'$. The open string between the 2 D-branes is coupled to a closed string external background NS $B$-field and to  a background abelian gauge connection $A$ with curvature $F=dA$ . Within the Type II theory one can have the following observable
\begin{equation}
\left \langle \int_{W} C_p (x) \cdot
\int_{W'} \exp \left ( \int_\Sigma B - i \int d\sigma A_I(X) \partial_\tau X^I \right) C_{p'} (y) \right
\rangle,
\end{equation}
where $X$ is the embedding of the world-sheet $\Sigma$ into the target space manifold $M$, $\tau$ and $\sigma$ are the open string world-sheet coordinates and $I=0, \cdots, 9$. Of course in addition to the observables $\int_W C_p$ we have that the term $\int d\sigma A_I(X) \partial_\tau X^I$ is invariant under gauge transformations of the background connection $A \to A + d \lambda$. There is also a combined invariance under changes of $B$ and $A$ as follows: $\delta B_{IJ}= \partial_I \Lambda_J - \partial_J \Lambda_I$ and $\delta A_I = - \Lambda_I$. (Compare with the {\it Wilson surface} introduced by Cattaneo and Rossi in Ref. \cite{Cattaneo:2002tk}.)

The corresponding asymptotic version is given by gauge invariant quantity
\begin{equation}
\left\langle \int_M i_{{\bf X}_p}(C_p) \mu_{{\bf X}_p}(x) \cdot  \int_{W'} \exp \left( \int_\Sigma B - i \int d\sigma A_I(X) \partial_\tau X^I \right) i_{{\bf Y}_{p'}} (C_{p'})\mu_{{\bf Y}_{p'}}(y) \right \rangle.
\end{equation}
Just as in the case of $BF$ theory it is required that the above correlation function be gauge invariant. We can see that that is precisely the case if one of D$p$'s is and anti-D-brane. The reason is as follows: it is well known that from the CS-action (\ref{cs-action}) that the RR-fields carries also $U(1)$-charge. Thus the field $C_p$ is charged under $U(1)$ as $1$ while the $C_{p'}$ transform as $-1$. this implies that the whole observable and its correlation function will be gauge invariant. In a similar spirit to the $BF$ theories this quantity would compute some linking number of couple of non-intersecting flows two D-branes and anti-D-branes with a non-trivial phase given by the non-trivial class $ \oint {B \over 2 \pi}$.

It is known that the true framework to deal D-branes is not cohomology but K-theory. The description of D-branes in the set up of dynamical systems lead to the possibility of extending the asymptotic cycles as homology classes to K-homology cycles in K-theory.

\section{Final Remarks}

In the present paper we pursue the idea of the implementation of
the procedure followed in Ref. \cite{VV} for Jones-Witten
invariants, to compute link invariants for flows in higher
dimensions. The relevant invariants of interest were
elucidated in Refs. \cite{Horowitz:1989km,Blau:1989bq}. We were able to obtain the higher-dimensional
generalization of the  asymptotic linking numbers for one flow (\ref{thopf1}) and for two flows (\ref{hopf-helicity}). We also obtain the generalized linking numbers in the non-abelian case for one flow (\ref{conn1}) and two flows (\ref{conn2}). Therefore for all these mentioned cases we were able to
associate a link invariant in higher dimensions to a flow (or flows). We calculate
the linking number for a flow (foliation), with this we were able to find the higher dimensional $n=2p+1$ generalization of asymptotic Hopf invariant and consequently of the Jones-Witten invariants for flows (\ref{highJW}) considered in \cite{VV} for the three dimensional case. Finally some speculations about a way to incorporate asymptotic cycles was discussed in Sec. (6) and we found that one condition to find gauge invariant correlation functions of observables constructed with RR-fields, $B$-fields and gauge fields impose that the flows associated to the D-branes correspond to a pair D$p$-$\overline{\rm D}p'$-brane (pair brane-antibrane). Rather than cohomology, RR-fields take values in $K$-theory, thus it would be interesting to carry over the construction of asymptotic cycles to $K$-homology. Some of this work is in progress.

To find these invariants we modify only the observables which have the information of the flow. These observables were constructed with the ideas of asymptotic cycles, geometric currents and foliations introduced by Sullivan \cite{S} and Schwarzman \cite{SH}.  Thus we take the asymptotic observables and the path integral give the asymptotic linking number a (Hopf invariant or helicity). Other invariants of knots and links in the context of Batalin-Vilkovisky (BV) quantization incorporating Wilson surfaces \cite{Cattaneo:2002tk} and string topology \cite{Cattaneo:2002rq} are our strong interest and will be considered in a future publication. Moreover a generalized asymptotic linking number (Sec. 4) would be extended by considering a $p$-form gauge potential \cite{Henneaux:1986ht}.

On the other hand, in other theories as the AdS/CFT
correspondence, there is also an underlying $BF$ theory of the
form $\int_Y B_{RR} \wedge dB_{NS}$ \cite{Witten:1998wy}. Thus the
observables of the theory are also susceptible to be extended as
there exist a flow determined by a vector field ${\bf X}$. Moreover, the
Hitchin functional $\int_{M_7} \Phi \wedge \star_7 \Phi$ is
defined on a 7 manifold $M_7$ of $G_2$-holonomy, which resembles a
$BF$ action. It is worth to mention that the partition function at
one-loop has been computed recently
\cite{Pestun:2005rp,deBoer:2007zu} in terms of the BV formalism by
obtaining the Ray-Singer torsion of $M_7$. It would be interesting
to find the observables of the theory and their asymptotic
counterparts. Some of the results on this subject will be reported
elsewhere.

\vskip 1truecm
\centerline{\bf Acknoledgements}
It is a pleasure to thank  B. Itz\'a, A. Mart\'{\i}nez-Merino, P. Paniagua and A. Pedroza for enlightening  discussions and useful suggestions.  This work of R.S. is supported in part by CONACyT  graduate fellowship.


\bibliography{octaviostrings}
\addcontentsline{toc}{section}{Bibliography}
\bibliographystyle{TitleAndArxiv}


\end{document}